\documentclass[aps,prl,reprint,twocolumn,superscriptaddress,showpacs]{revtex4-2}

\usepackage{graphicx}
\usepackage{mathrsfs}
\usepackage{bm}
\usepackage{amsmath}
\usepackage{dcolumn}
\usepackage{epstopdf}
\usepackage{dsfont}
\usepackage{amssymb}
\usepackage{tabularx}
\usepackage{array}
\usepackage{float}
\usepackage{color}
\usepackage{epstopdf}
\usepackage{mathrsfs}
\usepackage[colorlinks, linkcolor=blue,anchorcolor=blue,citecolor=blue,urlcolor=blue]{hyperref}
\usepackage{multirow}

\begin{document}
\title{Multiple types of unconventional quasiparticles in the chiral crystal CsBe$_2$F$_5$}

\author{Xin-Yue Kang}
\affiliation{School of Physics, Northwest University, Xi'an 710127, China}
\affiliation{Shaanxi Key Laboratory for Theoretical Physics Frontiers, Xi'an 710127, China}

\author{Jin-Yang Li}
\affiliation{School of Physics, Northwest University, Xi'an 710127, China}
\affiliation{Shaanxi Key Laboratory for Theoretical Physics Frontiers, Xi'an 710127, China}

\author{Si Li}
\email{sili@nwu.edu.cn}
\affiliation{School of Physics, Northwest University, Xi'an 710127, China}
\affiliation{Shaanxi Key Laboratory for Theoretical Physics Frontiers, Xi'an 710127, China}

\begin{abstract}
Unconventional topological quasiparticles have recently garnered significant attention in the realm of condensed matter physics. Here, based on first-principles calculations and symmetry analysis, we reveal the coexistence of multiple types of interesting unconventional topological quasiparticles in the phonon spectrum of the chiral crystal CsBe$_2$F$_5$. Specifically, we identified eight entangled phonon bands in CsBe$_2$F$_5$, which give rise to various unconventional topological quasiparticles, including the spin-1 Weyl point, the charge-2 Dirac point, the nodal surface, and the hourglass nodal loop. We demonstrate that these unconventional topological quasiparticles are protected by crystal symmetry. We show that there are two large Fermi arcs connecting projections of the bulk spin-1 Weyl point and charge-2 Dirac point on the (001) surface and across the entire surface Brillouin zone. Our work not only elucidates the intriguing topological properties of chiral crystals but also provides an excellent material platform for exploring the fascinating physics associated with multiple types of unconventional topological quasiparticles.

\end{abstract}
\maketitle
\textit{\textcolor{blue}{Introduction.}}
Crystalline materials offer an excellent platform for the study of topological quasiparticles in condensed matter physics~\cite{armitage2018weyl,zhang2019catalogue,vergniory2019complete,tang2019comprehensive,xu2020high,lv2021experimental,vergniory2022all}. Topological quasiparticles typically emerge around the band degeneracy points and have garnered significant attention over the past decade. The most well-known conventional topological quasiparticles include the twofold degenerate Weyl point~\cite{murakami2007phase,wan2011topological,armitage2018weyl} and fourfold degenerate Dirac point~\cite{young2012dirac,wang2012dirac,wang2013three}, which resemble relativistic Weyl and Dirac fermions and can simulate fascinating effects from high energy physics. However, the fundamental symmetry for condensed matter systems is the space group symmetry, which is a smaller subgroup of the Poincaré symmetry. This reduced set of constraints enables the emergence of a diverse range of unconventional quasiparticles beyond the Weyl and Dirac fermions. 
For instance, band degeneracies can lead to higher-dimensional structures such as nodal lines~\cite{burkov2011topological,weng2015topological,Chen2015,Mullen2015,Fang2015,Yu2015,Kim2015,Li2016,Bian2016,Huang2016,Yu2017,Li2017,Li2018nonsymmorphic} and nodal surfaces~\cite{zhong2016towards,liang2016node,bzduvsek2017robust,wu2018nodal}, as well as high-fold fermions (three-, six-, and eight-fold degenerate fermions)~\cite{weng2016topological,zhu2016triple,bradlyn2016beyond}, and higher-order fermions with quadratic and cubic dispersions~\cite{fang2012multi,yu2019quadratic}.

 The concept of topological quasiparticles was initially introduced within electronic systems, and subsequently, efforts were made to extend it to phonon systems~\cite{zhang2010topological,susstrunk2016classification,liu2017pseudospins,ji2017topological,liu2020topological}. Numerous materials featuring phonons with nodal points, nodal lines, and nodal surfaces have been theoretically predicted~\cite{zhang2018double,singh2018topological,xia2019symmetry,zhang2019phononic,liu2020symmetry,huang2020ideal,wang2020symmetry,xie2021three,liu2021ideal,li2021computation,xu2022catalogue}, and a portion of these predictions have been effectively validated through experimentation~\cite{zhang2019phononic,miao2018observation}.

Chiral crystals are materials characterized by a lattice structure that possesses a distinct handedness due to the absence of inversion, mirror, or roto-inversion symmetries. These crystals hold great promise across diverse fields, including biology, chemistry, medicine, optics, and materials science. Their chirality can profoundly influence optical, electrical, and mechanical properties, as well as interactions with other molecules.~\cite{wang2013emerging,yang2021chiral,ma2021recent}. Chiral crystals have demonstrated a wide array of captivating physical phenomena, encompassing skyrmions~\cite{bogdanov1994thermodynamically}, non-local and
non-reciprocal electron transport~\cite{rikken2001electrical,yoda2015current}, optical activity and magnetochiral dichroism~\cite{fasman2013circular}, ferroelectricity~\cite{mitamura2014spin,hu2020chiral}, the spintronics~\cite{yang2020spintronics}. More recently, significant attention has turned towards the topological properties of chiral crystals~\cite{shekhar2018chirality,chang2018topological}. Within these crystals, various novel topological chiral fermions have been discovered, such as Kramers–Weyl fermions and higher-fold degenerate  fermions~\cite{chang2018topological,sanchez2019topological,li2019chiral,xie2021kramers,bradlyn2016beyond,chang2017unconventional,tang2017multiple}.
\begin{figure}[htb]
	\includegraphics[width=8.5cm]{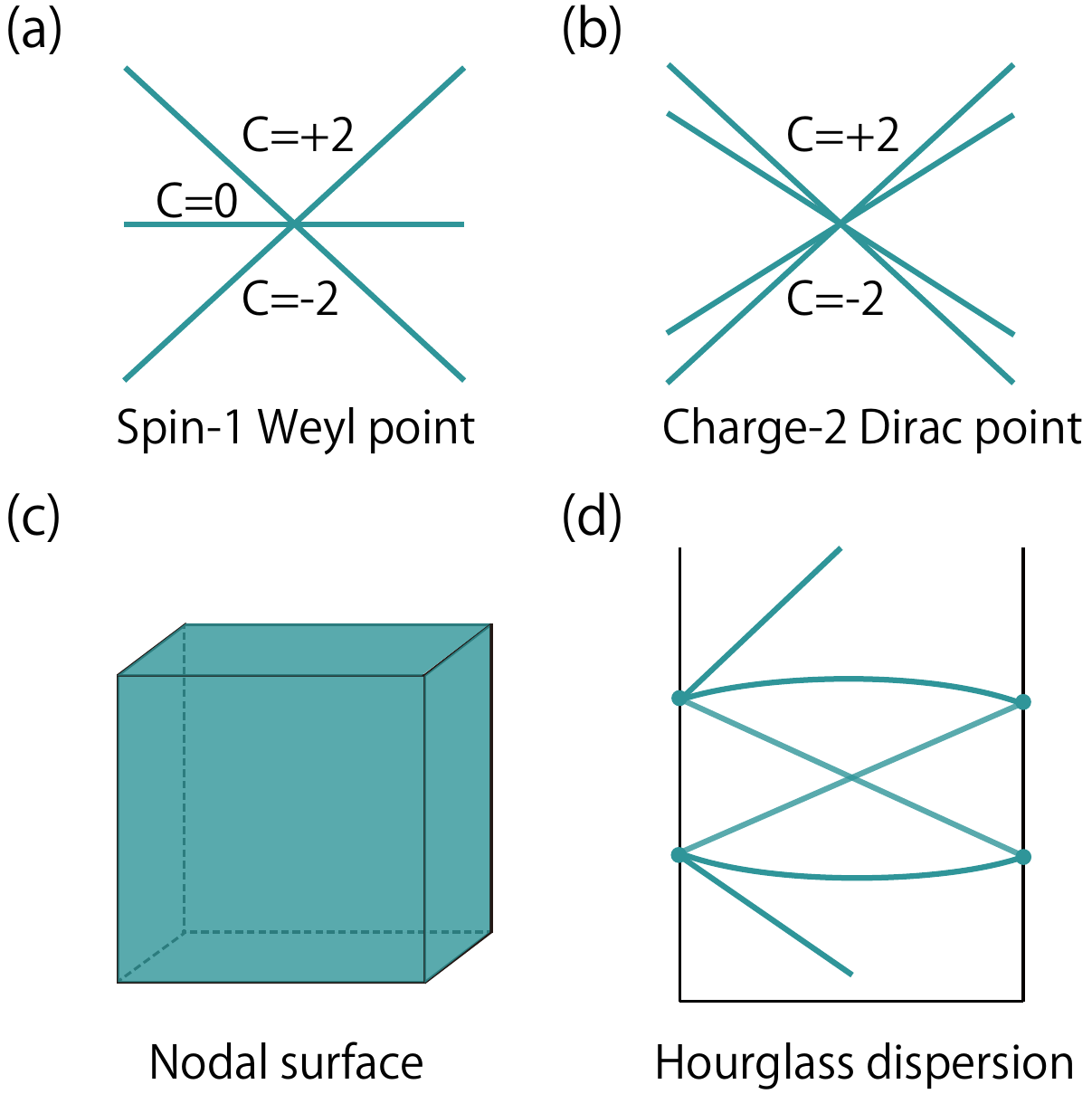}
	\caption{(a) Spin-1 Weyl point with Chern numbers
		of $\pm 2$, 0. (b) Charge-2 Dirac point with Chern numbers $\pm 2$. (c) Nodal surfaces locate on the Brillouin zone (BZ) boundary. (d) Hourglass dispersion emerges from the triple degenerate point and nodal surface.
		\label{fig1}}
\end{figure}

In this paper, based on first-principles calculations and symmetry analysis, we identify the chiral crystal CsBe$_2$F$_5$ material crystalizing in space group $P4_132$
(No.213) as an ideal material candidate
for studying the physical properties of multiple types of unconventional topological quasiparticles in its phonon spectrum. There exist eight phonon bands entangled together, forming several unconventional topological quasiparticles, such as the spin-1 Weyl points at the center of the Brillouin zone (BZ), the charge-2
Dirac points at the corner of the BZ, the nodal surface on the BZ boundary, as well as the hourglass nodal loop that emerges from the spin-1 Weyl points and nodal surface. The schematics of these unconventional topological quasiparticles are shown in Fig.~\ref{fig1}. The symmetry protection and interesting double-helicoid surface states of these unconventional topological quasiparticles are discussed.

\begin{figure}[htb]
	\includegraphics[width=8.5cm]{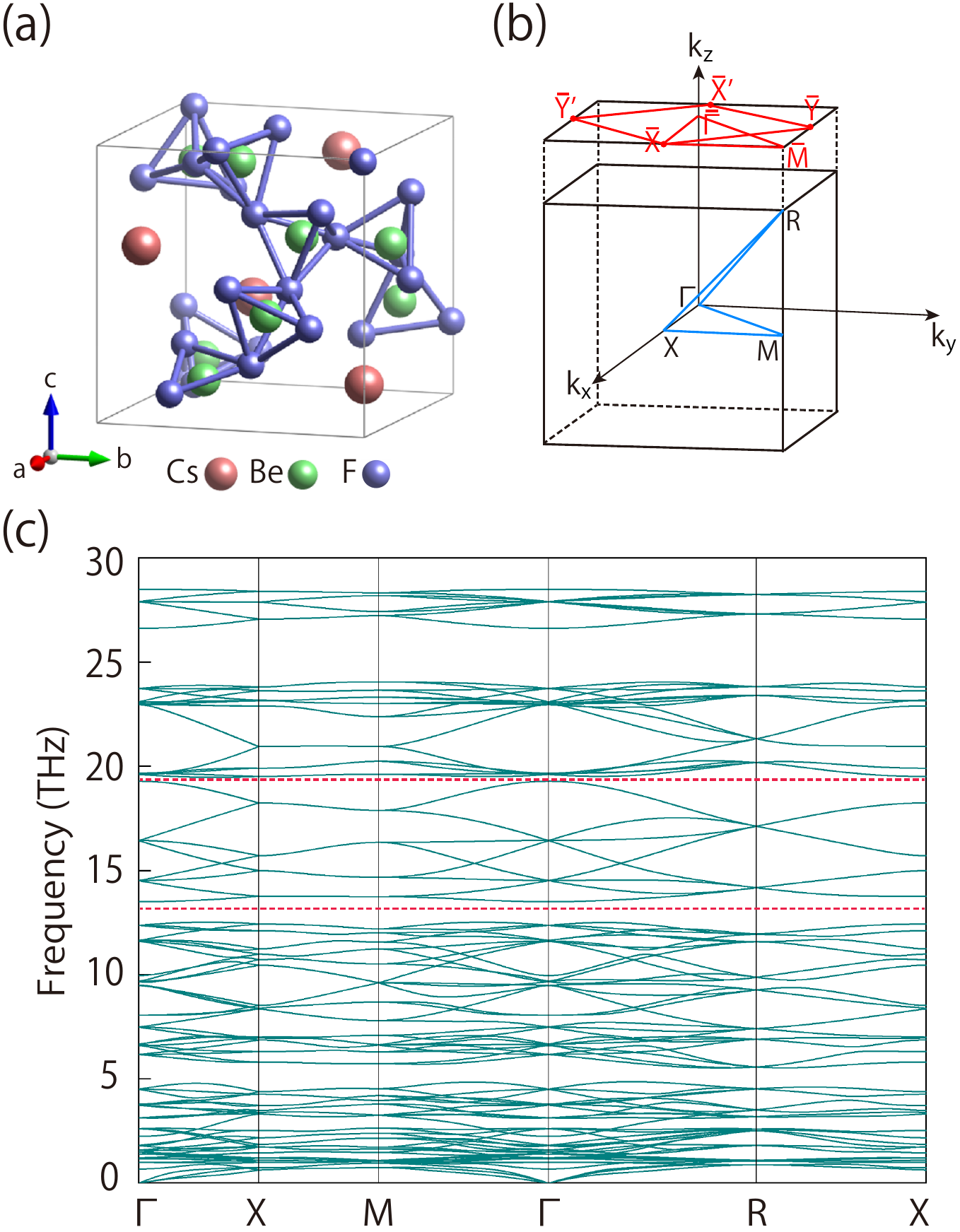}
	\caption{ (a) Crystal structure of the CsBe$_2$F$_5$. (b) Bulk BZ and
		the projected surface BZ of the (001) plane. The high-symmetry points are labeled. (c) Phonon dispersion of CsBe$_2$F$_5$ along the high-symmetry direction. The eight phonon bands (the 65th to the 72nd bands) are located between the two dashed red lines.
		\label{fig2}}
\end{figure}

\begin{figure}[htb]
	\includegraphics[width=8.5cm]{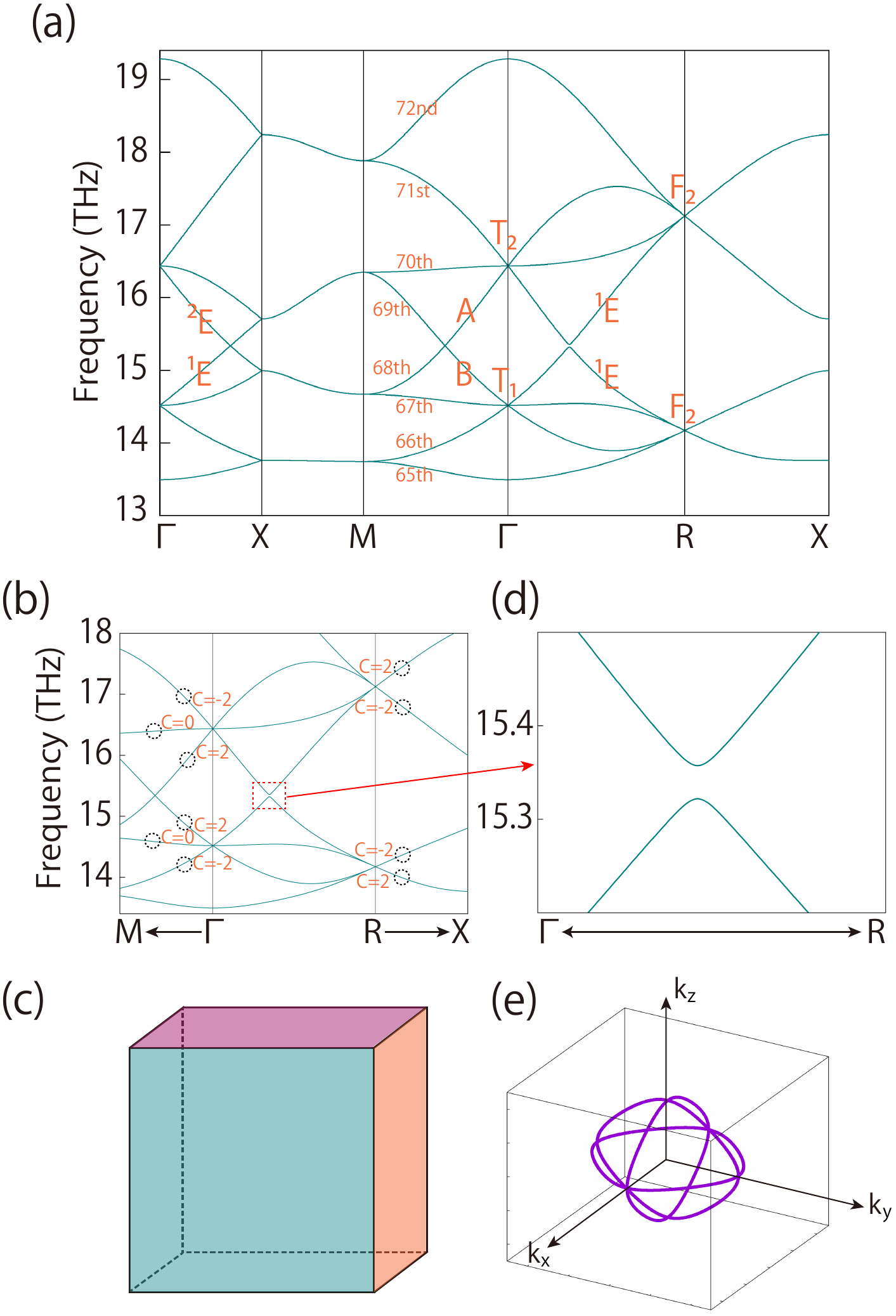}
	\caption{(a) The eight phonon bands of the CsBe$_2$F$_5$  (the 65th to the 72nd bands) between 13 and 19.4 THz. The irreducible representations (IRRs) are given. (b) The zoom-in of the phonon spectrum around the $\Gamma$ and $R$ points, and the calculated Chern numbers are given. The zoom-in of the phonon spectrum indicated by the box on the $\Gamma$-R path is shown in (d). (c) Schematic figure showing three nodal surfaces on the $k_i=\pi$ ($i=x,y,z$) planes. (e) The shape of the three hourglass nodal loops in the $k_i=0$ ($i=x,y,z$) planes obtained from first-principles calculations. 
		\label{fig3}}
\end{figure}

\textit{\textcolor{blue}{Lattice structure and phonon calculations.}}
The CsBe$_2$F$_5$ material has been experimentally synthesized and belongs to the chiral cubic lattice with space group $P4_132$ (No.213)~\cite{le1972structure}. The crystal structure of CsBe$_2$F$_5$ is shown in Fig.~\ref{fig2}(a). The unit cell comprises four molecules of CsBe$_2$F$_5$, with Cs and Be atoms occupying the 4a (0.125, 0.625, 0.875) and 8c (0.0998, 0.0998, 0.0998) Wyckoff positions, respectively, while F atoms occupy the 8c (0.7574, 0.7574, 0.7574) and 12d (0.125, 0.0334, 0.2834) Wyckoff positions. The atoms coordinates are measured
in units of the respective lattice parameters. In our calculation, the structure is
fully optimized (see Supplemental Material (SM)~\cite{SM} for computational details). The optimized lattice parameter for the  CsBe$_2$F$_5$
is 8.068 \AA , which is 
close to the experimental values, 7.936 \AA~\cite{le1972structure}. The bulk and (001)
surface BZs of CsBe$_2$F$_5$ are shown in Fig.~\ref{fig2}(b). The phonon spectrum obtained from first-principles calculations is shown in Fig.~\ref{fig2}(c). From Fig.~\ref{fig2}(c), one can observe eight phonon bands (the 65th to the 72nd bands) between 13 and 19.4 THz that are separated from other phonon bands and entangle together, as indicated by the two dashed red lines. The zoom-in image of these eight phonon bands is shown Fig.~\ref{fig3}(a), and the calculated irreducible representations (IRRs) for the phonon bands are also provided. 

As demonstrated in Fig.~\ref{fig3}(a), the eight phonon bands entangle together forming multiple types of unconventional quasiparticles, such as threefold degenerate points at $\Gamma$ point, fourfold degenerate Dirac points at $R$ point, twofold degenerate on the path X-M, and novel hourglass dispersion along paths $\Gamma$-X and M-$\Gamma$. In fact, the threefold degenerate points at $\Gamma$ point correspond to spin-1 Weyl points, the Dirac points at $R$ point correspond to charge-2 Dirac points, the twofold degeneracy along path X-M belong to a nodal surface on the $k_x = \pi$ plane [there are in total three nodal surfaces locate on $k_i = \pi$ $(i = x, y, z)$ planes], the neck point of the hourglass dispersion forms three hourglass nodal loops on the $k_i = 0$ $(i = x, y, z)$ planes. In the following sections, we will discuss these unconventional topological quasiparticles in detail.

\textit{\textcolor{blue}{Multiple Types of Unconventional Quasiparticles.}}
We first discuss the threefold degenerate points at the $\Gamma$ point. From Fig.~\ref{fig3}(a), we can observe two threefold degenerate points at the $\Gamma$ point, with three-dimensional IRRs $T_1$ and $T_2$, respectively. The calculated Chern numbers of the three bands that construct the threefold degenerate points are $\pm 2$ and 0, as shown in Fig.~\ref{fig3}(b). To further characterize the threefold degenerate topological quasiparticles, we construct a low-energy
effective $k \cdot p$ model based on the symmetry requirements. The generators of the little group at the $\Gamma$ point are $S_{2x}=\{ C_{2x}|\frac{1}{2},\frac{1}{2},0\}$, $S_{2y}=\{ C_{2y}|0,\frac{1}{2},\frac{1}{2}\}$, $S_3=\{C^{+}_{3,111}|0,0,0\}$, and  $S_{4x}=\{C_{4x}^+|\frac{1}{4},\frac{1}{4},\frac{3}{4}\}$. The three low-energy
states at the $\Gamma$ point belong to the $T_1$ (or $T_2$) representation. We use it as the basis and write the generators of representation matrices for $T_1$ (or $T_2$) of the $\Gamma$ point as follows:
\begin{equation}
	\begin{split}
		S_{2x}&=\left[\begin{array}{ccc}
			1 & 0 & 0 \\
			0 & -1 & 0 \\
			0 & 0 & -1 \\
		\end{array}\right],	
		S_{2y}=\left[\begin{array}{ccc}
			-1 & 0 & 0 \\
			0 & 1 & 0 \\
			0 & 0 & -1 \\
		\end{array}\right],\\	
		S_{3}&=\left[\begin{array}{ccc}
			0 & 0 & 1 \\
			1 & 0 & 0 \\
			0 & 1 & 0 \\
		\end{array}\right],
		S_{4x}=\left[\begin{array}{ccc}
			1 & 0 & 0 \\
			0 & 0 & -1 \\
			0 & 1 & 0 \\
		\end{array}\right]\\ 
	(S_{4x}&=-\left[\begin{array}{ccc}
		1 & 0 & 0 \\
		0 & 0 & -1 \\
		0 & 1 & 0 \\
	\end{array}\right] for ~ T_2).
	\end{split}
\end{equation}
It should be noted that at the $\Gamma$ point, the time-reversal $\mathcal{T}$ is preserved, given as $\mathcal{T}=\mathcal{K} \otimes I_{3\times3}$, where $\mathcal{K}$ is the complex conjugation operator. 
The generic form of the effective Hamiltonian of the threefold degenerate point at the $\Gamma$ point up to the first order of $\bf{k}$ is as following:

\begin{equation}
	H_W(k)=\alpha \cdot\left(\begin{array}{ccc}
		0 & i k_z & -i k_y \\
		-i k_z & 0 & i k_x \\
		i k_y & -i k_x & 0
	\end{array}\right)=\alpha \cdot \bf{k} \cdot \bf{S},
\end{equation}
where $\alpha$ is a real constant, and $S_i$ are the spin-1 matrix representations of the rotation generators. This model is just the spin-1 Weyl Hamiltonian. Hence, the threefold degenerate points at the $\Gamma$ point are termed as spin-1 Weyl point~\cite{zhang2018double}. The spin-1 Weyl point can be regarded as the monopole with topological charge +2. Through symmetry analysis, we have shown that the spin-1 Weyl point is protected by space group symmetry
  and the detailed analysis is presented in SM~\cite{SM}.

Then we investigate the fourfold degenerate Dirac points at the $R$ point of the BZ corner. As depicted in Fig.~\ref{fig3}(a), there are two fourfold degenerate points at the $R$ point, with both four-dimensional IRRs $F_2$. The calculated Chern numbers of the bands surrounding the Dirac points at $R$ are $\pm 2$, which indicates that they are charge-2 Dirac points, as shown in Fig.~\ref{fig3}(b). Furthermore, it was discovered that all bands at the $R$ point form charge-2 Dirac points. The symmetry analysis of the charge-2 Dirac points is presented in SM~\cite{SM}. Notably, the fourfold degeneracy at the $R$ point is independent of the materials and solely determined by the crystal space group symmetry. From Fig.~\ref{fig3}(b), we can also observe that the Chern numbers of the spin-1 Weyl and charge-2 Dirac points are opposite, and in
the BZ, they appear in pairs. Therefore, the total Chern number is zero, which is consistent with the no-go theorem~\cite{nielsen1981absence}. We also construct a low-energy effective $k \cdot p$ model for the charge-2 Dirac point. As mentioned above, the four low-energy
states at the $R$ point belong to the $F_2$ representation. We use it as the basis and write the generators of representation matrices for $F_2$ of the $R$ point as follows:
\begin{equation}
	 \resizebox{1.0\hsize}{!}
	{$\begin{split}
		S_{2x}&=\frac{1}{\sqrt{2}}\left[\begin{array}{cccc}
			-i & -1 & 0 & 0\\
			1 & i & 0 & 0 \\
			0 & 0 & i & -1 \\
			0 & 0 & 1 &-i \\
		\end{array}\right],	
		S_{2y}=\frac{1}{\sqrt{2}}\left[\begin{array}{cccc}
			i & -1 & 0 & 0\\
			1 & -i & 0 & 0 \\
			0 & 0 & -i & -1 \\
			0 & 0 & 1 & i \\
		\end{array}\right],	\\
		S_{3}&=\frac{1}{4}\left[\begin{array}{cccc}
			1& \sqrt{2}+i & \sqrt{6}+\sqrt{3}i &-{\sqrt{3}} \\
		 -\sqrt{2}+i & 1 & \sqrt{3} & \sqrt{6}-\sqrt{3}i \\
			 \sqrt{6}-\sqrt{3}i & -\sqrt{3} & 1 & \sqrt{2}-i \\
		      \sqrt{3} & \sqrt{6}+\sqrt{3}i & -\sqrt{2}-i & 1	\\
		\end{array}\right],\\
    	S_{4x}&=\frac{1}{4}\left[\begin{array}{cccc}
		1-\sqrt{2}i & -i & -\sqrt{3}i & \sqrt{6}i-\sqrt{3} \\
		i& -1-\sqrt{2}i &-\sqrt{3}-\sqrt{6}i &-\sqrt{3}i \\
		\sqrt{3}i &-\sqrt{3}-\sqrt{6}i & 1+ \sqrt{2}i & i \\
		-\sqrt{3}+\sqrt{6}i & \sqrt{3} i & -i & -1+\sqrt{2}i
	\end{array}\right]
	\end{split}$}
\end{equation}
At $R$ point, the time-reversal $\mathcal{T}$ is preserved, given as $\mathcal{T}=\mathcal{K} \otimes I_{4\times4}$, where $\mathcal{K}$ is the complex conjugation operator. The generic form of the effective Hamiltonian of the fourfold degenerate point at $R$ point up to the first order of $\bf{k}$ is as following:
\begin{equation}
H_D(k)=\left(
\begin{array}{cccc}
    \eta_1 & \eta_5 & \eta_3 & \eta_4 \\
	\eta_5^{\dagger} & -\eta_1 & \eta_4 & \eta_3 \\
	\eta_3^{\dagger} & \eta_4^{\dagger} & \eta_2 & \eta_6 \\
	\eta_4^{\dagger} & \eta_3^{\dagger} & \eta_6^{\dagger}	& -\eta_2\\
\end{array}
\right)
\end{equation}
where 
\begin{equation}
		 \resizebox{1.0\hsize}{!}
{$\begin{split}
	&\eta_1=\frac{\sqrt{2}}{4} \left(C_{1}-2 C_{2}\right) (k_x-k_y),~ \eta_2=\frac{\sqrt{2}}{4} \left(C_{1}+2 C_{2}\right) (k_x-k_y)\\
	&\eta_3=-i \frac{1}{2} \sqrt{\frac{3}{2}} C_{1} (k_x-k_y),~ \eta_4=-\frac{1}{2} \sqrt{\frac{3}{2}} C_{1}(k_x+k_y)\\
	&\eta_5= -\frac{C_{1}}{4} \left(\sqrt{2} ik_x+\sqrt{2} ik_y+4 k_z\right)-2\sqrt{2}C_{2} \left(k_x+k_y+\sqrt{2}ik_z\right)\\
	&\eta_6= \frac{C_{1}}{4} \left(\sqrt{2} ik_x+\sqrt{2} ik_y-4 k_z\right)+ \frac{C_{2}}{2}\left(\sqrt{2}ik_x+\sqrt{2}ik_y+2 k_z\right)
\end{split}$}
\end{equation}
and $C_1,C_2$ are real parameters. After a unitary transformation,
\begin{equation}
	H_D(k)=\beta\left(\begin{array}{cc}
		\bf{k} \cdot \bf{\sigma} & 0 \\
		 0 & \bf{k} \cdot \bf{\sigma} \\
	\end{array}\right)
\end{equation}
where $\beta$ is a nonzero constant. This model is the direct sum of two identical spin-1/2 Weyl points, and is also referred to as the charge-2 Dirac point.

Next, we analyze the twofold degeneracy bands along the path X-M. Figures~\ref{fig2}(c) and ~\ref{fig3}(a) clearly show that all of the phonon bands on the X-M path are twofold degenerate. In fact, through symmetry analysis and first-principles calculations, we reveal that all of the bands on the $k_x=\pi$ plane are twofold degenerate, and form the nodal surface, as schematically shown in the green color in Fig.~\ref{fig3}(c). Furthermore, we find another two nodal surfaces on the $k_y=\pi$ and $k_z=\pi$ planes, as shown in the orange color and crimson color in  Fig.~\ref{fig3}(c), respectively. Therefore, there are in a total of three nodal surfaces on the BZ boundary. The three nodal surfaces are protected by the combination of time-reversal symmetry $\mathcal{T}$ and two-fold screw rotation symmetry $S_{2i} (i = x, y, z)$. We take the nodal surface on the $k_x=\pi$ plane as an example and perform the symmetry analysis. Any point on
the $k_x=\pi$ plane is invariant under $TS_{2x}$. In the absence of spin-orbit coupling (SOC), we have $(TS_{2x})^{2}=e^{-ik_{x}}=-1$ on the $k_x=\pi$ plane. It leads to the Kramer-like degeneracy, and form the nodal surface on this plane. The nodal surfaces on $k_y=\pi$ and $k_z=\pi$ planes are protected by the $TS_{2y}$ and $TS_{2z}$,respectively, and can be analyzed similarly.  

We are now moving on to analyze the band crossings on the $\Gamma$-X and M-$\Gamma$ paths. As shown in Fig~\ref{fig3}(a), we can observe that the four phonon bands (the 67th to the 70th bands) cross, forming hourglass dispersions along the $\Gamma$-X and M-$\Gamma$ paths. It should be noted that there is a small gap for the 68th and 69th bands along $\Gamma$-R, as shown in Fig~\ref{fig3}(d). We also calculated the IRRs for the 68th and 69th bands along the paths $\Gamma$-X, M-$\Gamma$, and $\Gamma$-R, as shown in Fig~\ref{fig3}(a). One can realize that 68th and 69th bands have different
representations $^1E$ and $^2E$ along the path $\Gamma$-X, and $B$ and $A$ along the path M-$\Gamma$, while they have the same representations $^1E$ and $^1E$ along the path $\Gamma$-R. As a result, they can exhibit band crossing along paths $\Gamma$-X and M-$\Gamma$, and exhibit avoided crossing or anticrossing along path $\Gamma$-R.   Actually, the neck point (the band crossing point formed by 68th and 69th bands) of the hourglass are not isolated, and a careful scan shows
that the neck point are located on a nodal loop in the $k_z = 0$ plane around the $\Gamma$ point, forming the Weyl hourglass nodal loop. Furthermore, due to the symmetry operators $\{C^{+}_{4x}|\frac{1}{4},\frac{1}{4},\frac{3}{4}\}$ and $\{C^{+}_{4y}|\frac{3}{4},\frac{1}{4},\frac{1}{4}\}$, there should be another two Weyl hourglass nodal loops locate on the $k_y = 0$ plane and $k_x = 0$ plane. This is also certified by our DFT calculation, as shown in Fig~\ref{fig3}(e). These hourglass nodal loops are protected by the $\mathcal{PT}$ symmetry, and carry a Berry phase of $\pi$. It should be mentioned that the hourglass dispersions here emerge from the triple degenerate points at $\Gamma$ and the nodal surface on the $k_i=\pi$ ($i=x,y,z$) planes. They differ from previously reported hourglass fermions, which arise from either twofold degenerate points (in the case of Weyl hourglass fermions)~\cite{wang2016hourglass} or fourfold degenerate points (in the case of Dirac hourglass fermions)~\cite{Li2018nonsymmorphic}.

\begin{figure}[htb]
	\includegraphics[width=8.7cm]{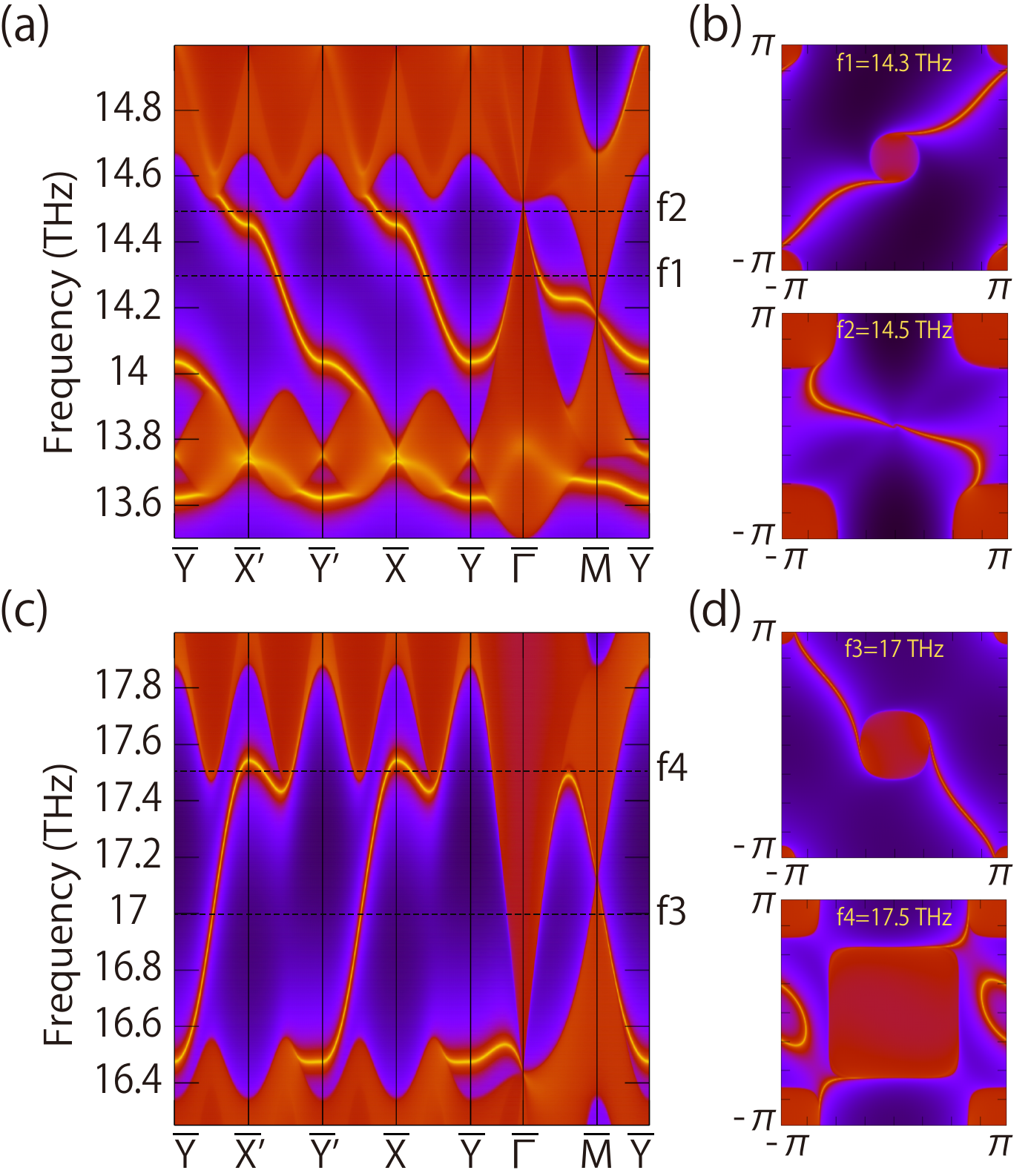}
	\caption{ Double-helicoid surface states for the (001) surface for the spin-1 Weyl points and charge-2 Dirac points along
		high-symmetry directions. (a) The surface states from 13.5 THz to 15 THz. (b) The corresponding Fermi arcs for 14.3 THz and 14.5 THz. (c) The surface states from 16.25 THz to 18 THz. (d) The corresponding Fermi arcs for 17 THz and 17.5 THz.
		\label{fig4}}
\end{figure}

Finally, we constructed an eight-band tight-binding model that accurately reproduces all the band crossings in CsBe$_2$F$_5$. The detailed information regarding the model and band structure is shown in SM~\cite{SM}.

\textit{\textcolor{blue}{Double-helicoid surface states.}}
Topological quasiparticles are typically characterized by the presence of surface states. The spectra of the (001) surface for the spin-1 Weyl point and charge-2 Dirac point of CsBe$_2$F$_5$ are shown in Fig.~\ref{fig4}. One can see that two large topological surface states emerge from $\bar{\Gamma}$ and $\bar{M}$. Here, the $\bar{\Gamma}$ point represents the projection of the spin-1 Weyl point at $\Gamma$ and $\bar{M}$ represents the projection of the charge-2 Dirac point at $R$. Additionally, the large Fermi arcs connecting the point $\bar{\Gamma}$ to the point $\bar{M}$ and across the entire surface BZ, as shown in Figs.~\ref{fig4}(b) and~\ref{fig4}(d). Previous work has shown that the Weyl surface states are equivalent to a helicoid~\cite{fang2016topological}. In the present case, since the surface states arise from the spin-1 Weyl point and charge-2 Dirac point, two surface sheets encircle these points, which would give rise to the double-helicoid surface states, as shown in Fig.~\ref{fig4}.

\textit{\textcolor{blue}{Discussion and Conclusion.}}
We have identified eight entangled phonon bands in CsBe$_2$F$_5$, leading to the emergence of diverse and unconventional topological quasiparticles. It is worth noting that in our recent work, we demonstrated that a band complex can have no upper bound for certain space groups, and find that an accordion-type band complex with $N_C=8$ (where $N_C$ denotes the number of bands in a complex) can be realized in the phonon spectra of the crystals AuCl and AuBr~\cite{li2023upper}. Here, the eight phonon bands in CsBe$_2$F$_5$ entangle together and form two distinct types of band complexes with $N_C=8$: one along the $\Gamma-X$ path and the other along the $M-\Gamma$ path. The spin-1 Weyl and charge-2 Dirac phonons can also appear in other space groups, such as space group $P2_13$ (No. 198)~\cite{zhang2018double,zhong2022material}. Additionally, it should be noted that the charge-3 and charge-4 Weyl phonons have recently been predicted~\cite{wang2022hourglass,liu2021charge,xiao2023realization}. 

In conclusion, our first-principles calculations and symmetry analysis reveal that the phonon spectrum of the chiral crystal CsBe$_2$F$_5$ harbors several types of unconventional topological quasiparticles. The entanglement of eight phonon bands gives rise to the spin-1 Weyl point, the charge-2 Dirac point, the nodal surface, and the novel hourglass nodal loop, all of which are protected by crystal symmetry. We have also shown that the surface states of CsBe$_2$F$_5$ exhibit intriguing properties, such as large Fermi arcs and double-helicoid surface states. In experiments, techniques such as neutron scattering~\cite{delaire2015heavy} and x-ray scattering~\cite{mohr2007phonon} can be used to probe bulk phonons, while high-resolution electron energy loss spectroscopy~\cite{zhu2015high}, helium scattering~\cite{harten1987surface}, and THz spectroscopy~\cite{wu2016quantized,wu2015high} are suitable for investigating surface phonons.

\bigskip
\begin{acknowledgements}
	This work is supported by the NSF of China (Grant No. 12204378).
\end{acknowledgements}

\bibliography{CBF_ref}

\end{document}